# New option for solving the climatic problems with non-thermal laser driven boron fusion and ultrahigh magnetic fields

Heinrich Hora[#]
Department of Theoretical Physics, University of NSW, Sydney, Australia
[#]h.hora@unsw.edu.au,

*In contrast to the broad stream of sustainable developments on fusion energy, new aspects are developed now by applying ultra-short, ultra-powerful laser pulses in a plasma-block ignition scheme by avoiding the well known difficulties of thermal-pressure instabilities and losses through using electrodynamic non-thermal energy conversion. A further advantage is given by the new 10 kilotesla magnetic fields for fusion of uncompressed proton-boron fuel which avoids problems of dangerous nuclear radiation.*

Of the many options for reducing the $CO_2$ emissions into the atmosphere down to the permitted level of 1960, one is nuclear fusion. But this option of fusion energy is not primarily mentioned. It is burning hydrogen into helium, the energy source of the sun and of the myriads of stars. To gain nuclear fusion energy in a power plant is, despite an enormous research effort over the last 60 years, a very complicated project. An economical power plant confining the reaction of plasmas of dozens of million degrees by using magnetic fields is hardly feasible before the next 30 years. On the other hand, when the laser was discovered in 1960, it was hoped that a faster solution may be possible. Despite great research achievement, a solution is still ways off. The hopes for a solution of the climate problem by fusion are largely deferred.

A radical change in this situation, however, has been opened recently when it became possible (A) to produce extremely short and tremendously intense laser pulses and (B) to combine this with magnetic fields measured at thousands of times higher than known before, and measureed only few months ago in Japan. This should permit for the first time a long-known, but very difficult, fusion reaction of ordinary hydrogen nuclei, protons, with the 80% present isotope 11 in boron available as huge quantities in mountains. It is particularly important that this HB11-fusion completely avoids the known radiation problems of nuclear energy. This results in a clean, inexhaustible and low cost energy source for the first time.

Some minor regrouping of the present billions of dollars research efforts should - against all previous results - enable the creation of ideal power plants within 10 to 15 years. Even if many individual issues and details have to be worked out in experiments and in computer analyses, this new concept is based on research findings of recent years, based on long-time tested research results.

In nuclear fusion with lasers since 1960, it was initially thought to consider thermal processes, i.e., extreme rapid heating of targets with very high thermal pressures, compres-sion and thermal ignition of fusion reactions. So it was possible in 1991 in Osaka/Japan that polyethylene could be squeezed together by lasers to two thousand times the solid density. The use of fusion fuel has reached high reaction yields. Laser pulses of nanosecond (ns, billionths of a second) duration reached yields close to the breakthrough for the fusion reactor some months ago with the largest laser in the world in Livermore/ California.

In contrast to these thermal methods for igniting fusion reactions, non-thermal but much more efficient processes have been studied for a long time, where laser energy can be converted directly into mechanical motion of high density plasma blocks. This avoids complex thermal processes, radiation, instabilities and retarding heat-transitions from electrons on the pressure-generating plasma ions. This goes back to Kelvin's discovery of the ponderomotive force of 1845, where electrically uncharged materials (as opposed to electric Coulomb forces) can be moved when electric fields have a special geometry. So you can, for example, move living cells under a microscope, without electric charging, in a desired manner. According to Kelvin, this force had to be further developed in due course, by inclusion of Maxwell's stress tensor showing the quadratic equations, i.e. those showing non-linear characteristics between the mechanical force and the force quantities of electromagnetic fields. This was crucial and it is called a non-linear force.

Since 1966, in Garching near Munich, thermally unexplained measurements of the laser-plasma interaction led to the introduction of nonlinear forces [1] based on the optical properties of high-temperature plasmas generated by the lasers at irradiation of materials. This led to the discovery of ordinary and relativistic self-focusing and plasma motions. At this time the work of Chu [2], parallel to that of Bobin, was studying how laser pulses must be during laser irradiation on the frozen simplest fusion fuels deuterium (D) and tritium (T) (heavy and superheavy hydrogen) in order to generate a fusion flame. The result was devastating. To reach nuclear ignition it needed an energy flux density E* of more than 100 Milllion joules per square centimeter during the extremely short time of a picosecond (ps, a millionth of a millionth of a second). In 1972 that was a complete utopia and it was decided that one has to work with the thousand times longer laser pulses of a Nanosecond, as operated in Lawrence Livermore's main direction of fusion based on thermal properties.

Nevertheless, ps-laser pulses were at this time - at least in theory and numerically - not unknown. Plasma-hydrodynamic computer calculations with all thermal details and the additional nonlinear forces of laser interaction showed in 1978 how laser pulses of 1.5ps duration and even at then realistic intensities of $10^{18}$Watts/cm$^2$, could accelerate 20 wavelengths thick blocks of deuterium plasma to velocities of $10^9$cm/s . These were ultra high accelerations [3], of more than $10^{20}$cm/s$^2$, which were needed for the ignition, according to Chu [2].

Before an experimental confirmation of this ultra high acceleration could be found, the discovery of the CPR method (Chirped Pulse Amplification) by Gerard Mourou and co-workers in 1985 was necessary [4]. Until then, the increase in the laser power had developed only slowly but with CPR it became a very dramatic turning point: the development curve makes a sharp bend upward. Since then, the laser intensity has increased (without self-focusing) by a factor of 10 Millions. In the ps or even much shorter laser pulses, the 10 PW (Petawatt = $10^{15}$ watts) power has been generated. This is about a thousand times greater than all the power plants on earth are producing.

Sauerbrey has measured these ultra high accelerations of plasma blocks of $2\times10^{20}$cm/s$^2$ in 1996 at Göttingen [5] in a directly visible way with the Doppler shift of spectral lines. That acceleration was 100,000 times higher than ever previously measured in a laboratory. This was in exact accordance with the 1978 theoretically predicted values [6]. These plasma blocks could be used to initiate the fusion flame for igniting of uncompressed solid DT fusion fuel with picosecond laser pulses after Mourou [4] as needed by Chu [2], whose hydrodynamic calculations needed some upgrading by later known effects.

When summing up this result [7] it turned out that the use of proton-boron (HB11) as a fusion fuel instead of DT, arrives at the same thresholds for laser ignition. That was a big surprise, and only possible because of the ps-laser pulses by Mourou which needed for non-thermal direct conversion of laser energy into macroscopic motion of the plasma blocks in contrast to the thermal-compressive ignition with ns laser pulses. In addition these calculations were pessimistic by using only the binary reactions as for DT. In fact an extremely higher energy gain occurs with HB11 by a secondary reaction due to elastic collisions of the resulting alpha particles with boron nuclei. These are occurring after the primary reaction, to generate three other alpha particles by an avalanche process with much higher energy gains than with DT.

The detailed processes calculated by Chu [2] were always expected in planar geometry. For a fusion power plant, lateral losses must be considered. The simplest solution is the use of the spherical geometry. When we calculated this with spherically fixed dense fusion fuel for both DT and for HB11 fuel, the energy gains are in the range of 100 only, even with using laser pulses of Exawatt, 1000 times higher than Petawatt [8]. At present, only 10 PW pulses can be produced although the European Billion-Euro project IZEST-ICAN-ELI is aiming the Exawatt (10^18 watts) and more, which is desirable for many other research apart from nuclear fusion.

After this radically changing result (A), another new discovery (B) is to help here by the production of 10 kilotesla magnetic fields, which are a thousand times greater than previously possible, but only for a period of slightly more than a ns and in volumes of several cubic-milllimeters [9]. For HB11-fusion that is just enough. When the cylindrical fusion fuel is captured with parallel axial magnetic fields of this magnitude (Fig. 1), following results [10] have been calculated for HB11: A ps laser pulse of 30 kilojoules of energy (30 PW power) can produce energy in alpha particles (helium nuclei) of about one gigajoule (GJ). This energy can be electrostatically converted into electrical power with high efficiency and low heat losses as used by Kanngiesser's HVDC High Voltage Direct Current transmission systems. The discharge of 1.4 MV within one second refers to an elecgric current of 720 Amps. One GJ equals 277 kWh. Only a small part of this energy is needed for the operation of the laser pulses 1 and 2 in a reactor according Fig. 2. This could be utilised in a power station with a reaction rate of one per second [11]. After deduction of the running costs, such a power station could be run economically and with handsome profits - or reduced costs - for absolutely clean and inexhaustible energy.

Attention should be given to recent experiments with HB11 fusion by Korn et al [12], whose fusion gains were an astonishing 1000 times higher than before [13]. These gains were not too far away from the highest gains ever measured with the biggest lasers on earth using the neutron generating DT fuel. And these experiments [12] had not at all the optimized lasers with more than 100 times longer pulses and with less contrast, than needed with the picoseconds block ignition [7], nor using the ultrahigh magnetic fields [9] produced by another nanosecond laser beam.

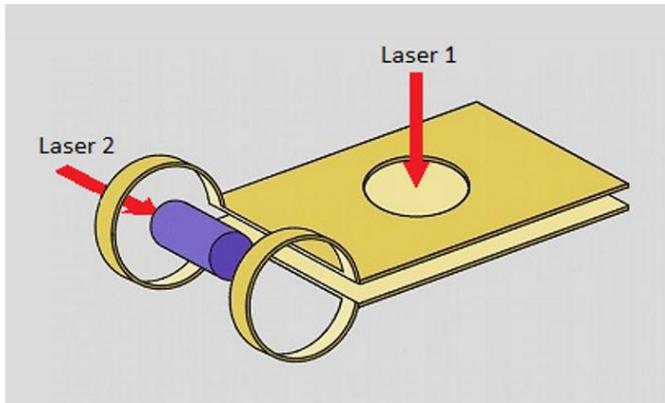

Figure 1. Generation of a 5 kilotesla magnetic field in the coils (Fujioka et al [9]) by firing a kilojoule nanosecond laser 1 into the hole between the plates. The HB11 fusion fuel is coaxially located in the coils and the block-ignition of the fusion flame is produced by the ps-30 kilojoule pulse from laser 2.

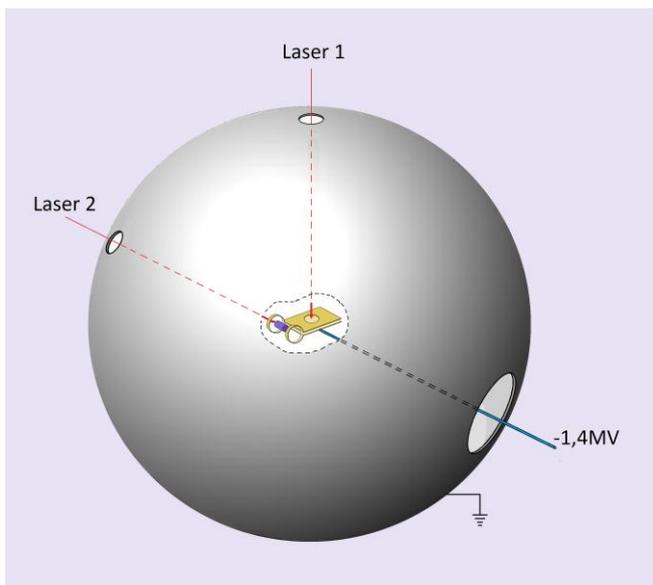

Figure 2. The HB11 fusion without a radioactive radiation problem is based on a block ignition [6][7] by a 30kJ-picosecond laser pulses 2 where the solid hydrogen-boron fuel in the cylindrical axis of the magnetic coil is trapped by 10 kilotesla field sustained for nanoseconds after generated by a nanosecond long laser pulse 1. The reaction is charged by -1.4 million volts against the wall of a sphere producing alpha particles (helium nuclei) of more than a gigajoule energy, of which a small part is needed for the operation of the lasers. One part of the gained costs of electricity is needed for the apparatus of the reaction and for the boron metal of the fuel [10] .

## 10 kilotesla magnetic field confinement combined with ultra-fast laser accelerated plasma blocks for initiating fusion flames*

Heinrich Hora[1], P. Lalousis[2], Shalom Eliezer[3,4], G. H. Miley[5], S. Moustaizis[6], G. Mourou[7]

1- Department of Theoretical Physics, University of New South Wales, Sydney 2052, Australia
2- Institute of Electronic Structure and Lasers FORTH, Heraklion, Greece
3- Institute of Nuclear Fusion, Polytechnic University of Madrid, Madrid, Spain
4- Soreq Research Center, Yavne, Israel
5- Dept. Nucl. Plasma & Radiol. Engineering University of Illinois, Urbana IL, USA,
6- Techn. Univ. Crete, Chania, Greece
7- LOA [Laboratoire dOptique Applique], ENSTA, Palaiseau Cedex, France

The measured magnetic fields of 5 Kilotesla [1] in combination with CPA generated picosecond laser pulses up to exawatt may open a new scheme of laser driven fusion with ultrahigh acceleration of plasma blocks [2]. The scheme with spherical irradiation of fusion fuel [3] has been modified to cylindrical geometry. Computations with 10 kilotesla magnetic fields for controlling losses show perfect confinement up to several ns if the initiation of the fusion flame in proton-11B fuel with ps laser pulses is used. The stopping power of alpha particles causes secondary reactions for increasing the fusion gains. Cylindrical geometry is then more favourable than the spherical case. The present results show that laser pulses of more than 30 petawatt are needed to produce GJ fusion energy. Support by circular polarization of the laser field interaction was studied and the high electric fields in the double layer for the ps laser initiation process [4].

*Supported by the Gordon Godfrey Fund, University of New South Wales, Sydney